\documentclass[]{article}

\usepackage[utf8]{inputenc} 
\usepackage[T1]{fontenc}    
\usepackage[sc]{mathpazo}   
\usepackage{hyperref}       
\usepackage{url}            
\usepackage{booktabs}       
\usepackage{amsfonts}       
\usepackage{microtype}      
\usepackage{graphicx}
\usepackage{comment}

\usepackage[english]{babel} 
\usepackage[hmarginratio=1:1,top=32mm,columnsep=20pt]{geometry} 
\usepackage[hang, small,labelfont=bf,up,textfont=it,up]{caption} 
\usepackage{booktabs} 
\usepackage{lettrine} 

\usepackage{abstract} 

\usepackage{titlesec} 
\titleformat{\section}[block]{\large\scshape\centering}{\thesection.}{1em}{} 
\titleformat{\subsection}[block]{\large}{\thesubsection.}{1em}{} 
\titleformat{\subsubsection}[block]{}{\thesubsubsection.}{1em}{} 

\usepackage{fancyhdr} 
\usepackage{titling} 
\usepackage{multicol}
\usepackage{wrapfig}
\usepackage{float}

\title{Digital Ariadne:\\ Citizen Empowerment for Epidemic Control}

\author{
  Lorenz Cuno Klopfenstein$^\ast$, Saverio Delpriori$^\ast$,\\
  Gian Marco Di~Francesco, Riccardo Maldini,\\
  Brendan Dominic Paolini, Alessandro Bogliolo$^\ast$\\
  \\
  DIGIT~srl, Urbino, Italy \\
  $\ast$~DiSPeA, University of Urbino, Italy
}

\begin{document}
\maketitle

\begin{abstract}
\noindent
The COVID-19 crisis represents the most dangerous threat to public health since the H1N1 influenza pandemic of~1918.
So far, the disease due to the SARS-CoV-2 virus has been countered with extreme measures at national level that attempt to suppress epidemic growth.
However, these approaches require quick adoption and enforcement in order to effectively curb virus spread, and may cause unprecedented socio-economic impact. 
A viable alternative to mass surveillance and rule enforcement is harnessing collective intelligence by means of citizen empowerment.
Mobile applications running on personal devices could significantly support this kind of approach by exploiting context/location awareness and data collection capabilities.
In particular, technology-assisted location and contact tracing, if broadly adopted, may help limit the spread of infectious diseases by raising end-user awareness and enabling the adoption of selective quarantine measures.
In this paper, we outline general requirements and design principles of personal applications for epidemic containment running on common smartphones, and we present a tool, called ``diAry'' or ``digital Ariadne'', based on voluntary location and Bluetooth tracking on personal devices, supporting a distributed query system that enables fully anonymous, privacy-preserving contact tracing.
We look forward to comments, feedback, and further discussion regarding contact tracing solutions for pandemic containment.

\end{abstract}

\noindent
\textbf{Keywords:} 
Citizen empowerment; Collective intelligence; Mobile app; Contact tracing; Epidemic control; Privacy; Anonymity; COVID-19.

\begin{multicols}{2}

\section{Introduction}

The novel coronavirus SARS-CoV-2 and its rapid spread have established a pandemic of global proportions over the course of the first months of~2020.
High fatality rates detected in the first affected regions are expected to be even higher in countries with an older population, low-income, or lack of suitable healthcare facilities~\cite{li_age-dependent_2020}.
In the absence of a viable vaccine, so far, the spreading of the disease due to the SARS-CoV-2 virus has been countered in many countries with countermeasures that attempt to suppress epidemic growth, thus avoiding to overwhelm the healthcare system with an unmanageable number of patients.
The reduction of contagion is achieved through a set of increasingly severe measures that limit personal freedom and entail strong socio-economic drawbacks, going well beyond mass gathering prohibition and case isolation. Social distancing rules, including school and university closure, household quarantine, internal and cross-border mobility constraints, and selective closure of non-essential productive and commercial activities, have brought many countries to complete lockdown~\cite{ferguson_report_2020}.

All these approaches require quick adoption and strict enforcement, in order to effectively curb virus spread in the short term.
As observed in the 1918~pandemic, there is a strong correlation between excess mortality and earliness of containment measures.
Containment interventions that are introduced too late or lifted too early were shown to have very limited effect~\cite{bootsma_effect_2007}.

In the long term, governments are required to trade off the adoption of dramatic full lockdown measures with more lax interventions.
For instance, progressively adopting temporary small-scale contagion suppression actions that aim at keeping the virus' \textit{reproduction number}, $R_0$, at a level that does not exceed the healthcare system's capacity.
Adaptive adoption of these kinds of containment policies at a regional level is expected to be effective even if enforced for shorter periods of time~\cite{ferguson_report_2020}.

The triggering of circumscribed quarantine measures can be directed through the widespread adoption of technological tools that allow tracing contacts and interactions with known cases of contagion~\cite{ferretti_quantifying_2020}.
Mobile apps running on personal smartphones are especially attractive as solutions because they enable immediate deployment on existing hardware and a quick response~\cite{hellewell_feasibility_2020}.
Several approaches of this kind have been proposed over the course of the last weeks, giving raise to a growing debate around privacy implications and potential risks of mass surveillance and stigmatization~\cite{abeler_support_2020,cho_contact_2020}, that have prompted authorities to provide recommendations and guidelines \cite{european_commission_commission_2020}, and big players to develop \textit{ad hoc} cross-platform protocols \cite{noauthor_apple_2020}.

In this paper, we suggest that these contact tracing tools should be designed to support end-user empowerment, as opposed to mass surveillance, granting citizens more data, awareness, and control, as envisioned by Nanni et al.~\cite{nanni_give_2020}.
In Section~\ref{sec:2req}, we outline the basic requirements and the founding principles on which they should be based.
In Section~\ref{sec:3diary}, we present a location/contact tracing solution, composed of a mobile app and a distributed query system, designed to meet these critical requirements.
The proposed system allows individuals to keep track of movements and contacts on their own private devices and to use local traces to select relevant notifications and alerts from health authorities, thus completely eschewing, by design, any risk of surveillance.

\section{Requirements and Design Principles}
\label{sec:2req}

Taking end-user empowerment as the founding principle, in this section we outline requirements and design principles that address both regulatory and technical issues.
Compliance with national and international regulation is essential to protect natural persons and their fundamental rights and freedoms, while technical requirements are mainly meant to reconcile dependability needs with the features of general-purpose personal devices, characterized by software fragmentation, hardware diversity, variety of non-exclusive usage modes, lack of calibration, limited resources, and untrained users.

\paragraph{a.~Collective intelligence.}
Systems based on the voluntary participation of individuals, performing a collective effort in their pursuit of a common goal, leverage a form of collective intelligence, which is the only alternative to mass surveillance and enforcement.
ICT~solutions should support and encourage such collaborative behaviors.

\paragraph{b.~Social responsibility.}
Individual participation in a collective effort towards a common goal is an act of social responsibility.
The technology adopted must make the social value of end-user's behavior clearly perceptible.

\paragraph{c.~Awareness and control.}
Technology is not infallible and systems may not always behave as expected.
Mobile apps should not induce end-users to simply rely on them.
Rather, they should empower end-users by granting them control and awareness of the data gathering process and by allowing them to browse their data and possibly add spontaneous notes.


\paragraph{d.~Privacy and anonymity by design.}
Protection of sensitive data, such as locations or health-related information, cannot rely exclusively on trust or security promises.
The system must be designed to keep user data private at all times, ideally storing them exclusively on the user's device, and to make identification impossible a~posteriori.

\paragraph{e.~Technology agnosticism.}
Contact tracing is a challenging task.
In spite of the many approaches that have been proposed, no single technology has proven to offer the ultimate solution.
For instance, location services have limited accuracy, especially indoor, while Bluetooth proximity does not reveal the exposure to indirect contagion (through infected surfaces).
The solution of choice should exploit all available technologies and be open to any improvement or integration.


\paragraph{f.~Effectiveness.}
The effectiveness of containment measures based on the voluntary adoption of a mobile app strongly depends on the percentage of the population, making proper use of that app.
Although this is always true, each solution has to be evaluated in different scenarios, including those well below the nominal critical mass of the target technology.
Two types of performance indicators have to be used, to measure the support that app can provide both to end-users tested/diagnosed positive to~COVID-19, willing to cooperate with health authorities, and to all other individuals possibly infected by them, who should take timely countermeasures.

\paragraph{g.~Interoperability.}
Interoperability must be pursued as much as possible, in order to reduce the critical mass requirements of each single system and to fully exploit their potential.
To this purpose, open standard protocols should be preferred to closed \textit{ad hoc} ones, cooperation among institutions must be technically supported, and integration with synergistic healthcare systems must be enabled.

\paragraph{h.~Openness of source code.}
Open-source access to all system components is the key to speed up development, ensure continuous improvement, and guarantee coherence between specification and implementation.
Transparency is essential both for end-users and for health authorities possibly adopting the solution.

\paragraph{i.~Openness of statistical data.}
Statistical data can provide valuable information to evaluate the effectiveness of epidemic containment, to monitor contagion, and to drive timely decisions.
All statistical information that can be provided by end-users on a voluntary basis, without jeopardizing their privacy and anonymity, is worth being gathered and made available as an open dataset.
Open data enables study and research without providing questionable competitive advantages to any player.

\paragraph{j.~Avoidance of false alarms.}
The ultimate goal of contact tracing systems is to reach susceptible or asymptomatic individuals who are considered to be the target of specific measures (e.g., quarantining or testing) according to the containment policies adopted.
The solution adopted must minimize unneeded alarms that can overwhelm the healthcare system and spread panic.

\paragraph{k.~Avoidance of surveillance and stigmatization.}
The fear of being watched and stigmatized may hinder the acceptance of contact tracing technologies.
Ideal solutions should prevent mass surveillance and stigmatization by design, by keeping end-user traces in their own devices and by making it impossible for end-users to identify the source of contagion.

\paragraph{l.~Scalability.}
The higher is the adoption rate, the more effective the solution is.
Hence, scalability is a key requirement.
Since the target devices, i.e., smartphones, have their own storage, computation, sensing, and communication resources, scalability can be inherently achieved by exploiting local resources as much as possible without triggering any network effect.

\section{Digital Ariadne}
\label{sec:3diary}

\textit{Digital Ariadne} or `diAry' is a privacy-preserving open-source tool, developed by DIGIT~srl and the University of Urbino, that allows users to trace their movements and contacts, while also allowing governments or healthcare agencies to rapidly direct their epidemic containment efforts, in a way that aligns with the principles outlined above.

The system is composed of: a \textbf{mobile application}, that is voluntarily installed by users on their smartphones, keeping track of their locations through the device's GPS sensor and interactions with other users through Bluetooth radio beacons, a \textbf{privacy-aware reward system}, which incentivizes app usage while collecting anonymous usage information to feed an open data set, and a \textbf{distributed query system} that allows recognized public authorities to selectively and anonymously notify users about possible contagion sources.
The mobile app works in background, with careful usage of battery and storage and without impairing the functioning of the personal devices. Nonetheless, it provides a rich user interface to make end-users fully responsible and aware of their own contribution to epidemic containment.

Source code of the mobile applications and the back-end service, both currently in active development, is available on GitHub\footnote{Source code: \url{https://github.com/digit-srl/diAry-apps} and \url{https://github.com/digit-srl/diAry-backend}.}.

\subsection{Mobile app}


The \textit{Digital Ariadne} mobile application is developed using the \textit{Flutter} framework for Apple~iOS and Google~Android.
A combination of movement detection with the built-in accelerometers, activity recognition, data from GPS sensors, and Bluetooth Low Energy~(BLE) transmission is used to adaptively track the user's movements and interactions without negatively impacting battery and storage capacity of the device. Traces are collected autonomously by a background service launched by the application, but end-users can decide at any time to interact with the app to browse stored data, to force sampling, to mark known locations or the add notes.

\begin{figure}[H]
 \centering
 \includegraphics[width=0.6\columnwidth]{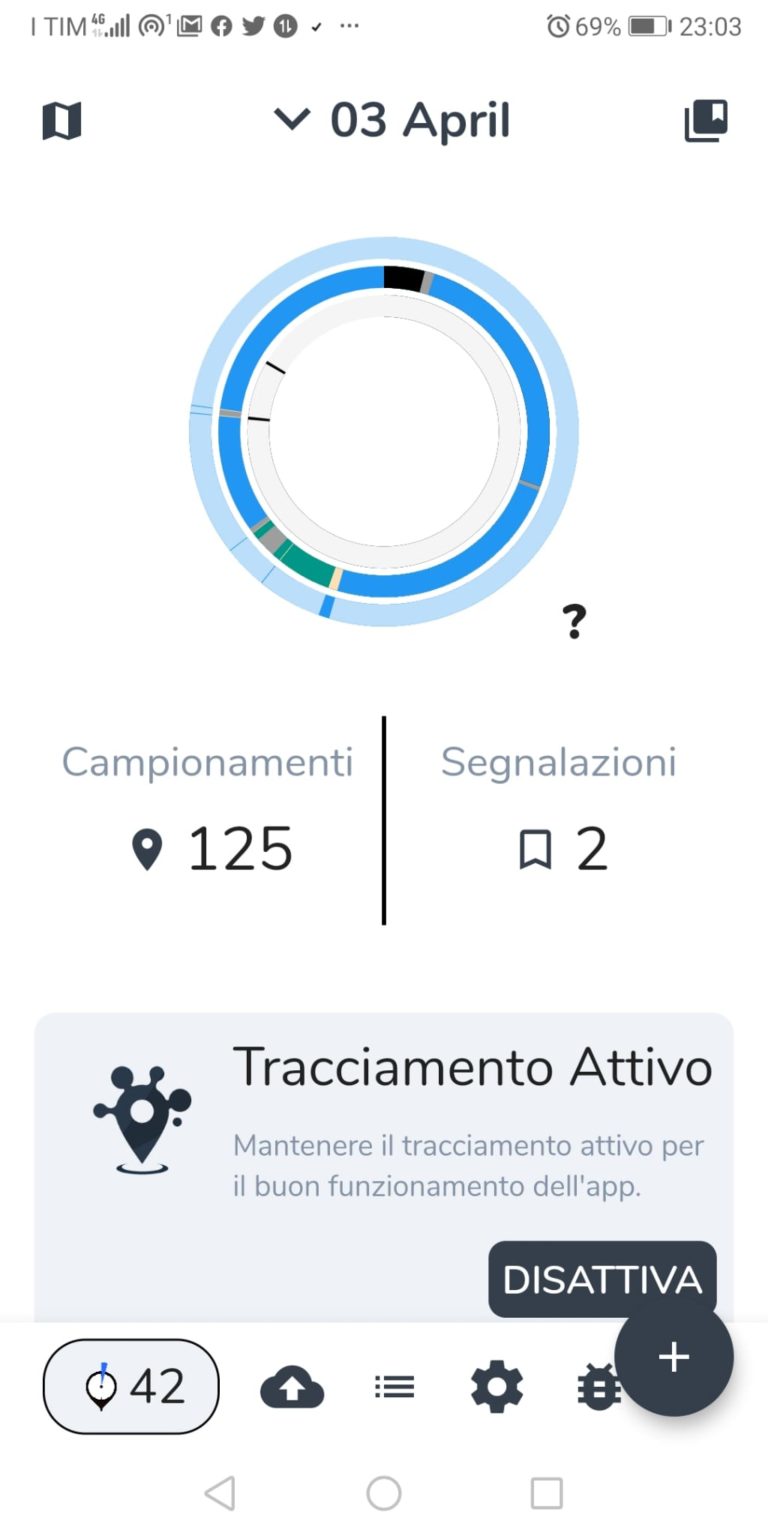}
 \caption{Main screen of the application.}
 \label{fig:mobile-screen-1}
\end{figure}

Tracking status is displayed on the main app screen, as shown in Figure~\ref{fig:mobile-screen-1}.
Three concentric circles, representing the 24 hours of the current day, show the detected movements, the amount of time spent in known locations, and the notes added by the end-user.

This information is stored for a maximum duration of 30~days on the device.
The app approximately requires 1~MB of data to store information about a day of full tracking (this requirement may slightly increase in case of frequent movement).

\subsubsection{Location tracking}

The app, once activated by the user, starts tracking the device's location and records detected positions and movements.
Location tracking is adaptive, based on the user's activity and speed, in order to provide sufficient precision in the case of movement and low battery consumption otherwise.

\begin{figure}[H]
 \centering
 \includegraphics[width=0.6\columnwidth]{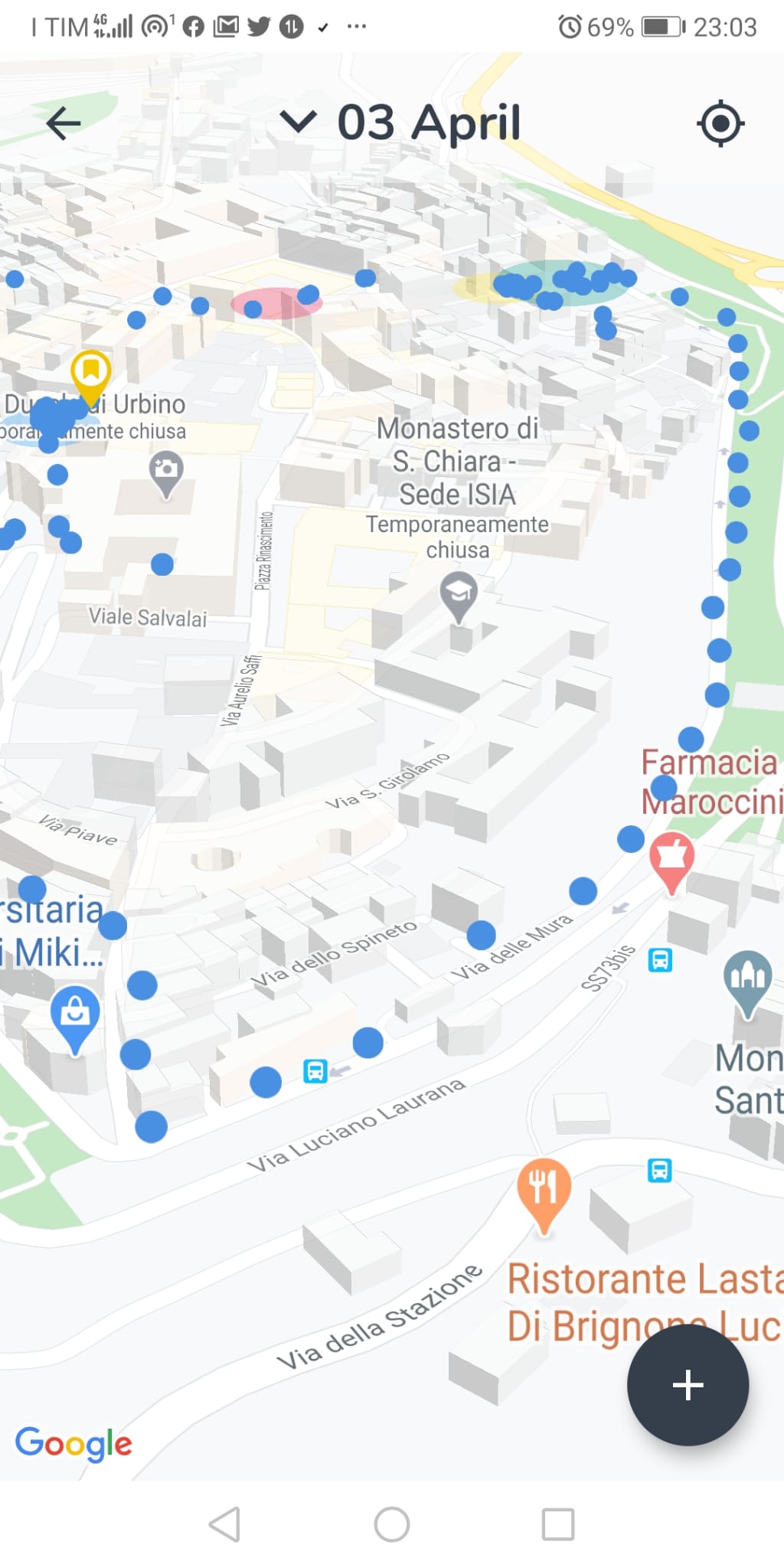}
 \caption{Map view of tracked locations.}
 \label{fig:mobile-screen-2}
\end{figure}

The user may voluntarily mark known locations through the app interface, thus allowing to specify places such as home, workplace, school, or any other locations.
Access to and departure from these locations are detected through geofencing and allow the user to have a quick overview of his or her movements throughout the day.

Also, the user may decide to add notes to a specific location and time of the day, in order to remember specific events or situations that may be relevant for contact tracing purposes.
User movements, known locations, and notes are shown on an interactive map like in Figure~\ref{fig:mobile-screen-2}.

\subsubsection{Interaction tracking}

The diAry app makes use of the \textit{Temporary Contact Numbers}~(TCN) contact tracing protocol in order to broadcast randomly-generated and anonymous identifiers, which are updated every few minutes~\cite{tcn_coalition_global_2020}.
All installations of the app keep a fully-local log of every identifier that has been broadcast and every identifiers that was received from other devices making use of the same TCN~protocol.
This data, which expires together with location data after 30~days, never leaves the phone and is not linked to private user information.

\subsection{Statistical open data set}

The \textit{Digital Ariadne} system makes use of a server-side component that is used to collect daily usage statistics in an anonymous fashion.
Users are never required to identify themselves and no user-identifying information is transmitted at any point.
Communication between mobile applications and the back-end makes use of secure connections using widely-adopted standards (HTTPS with optional certificate pinning).

An anonymous installation ID is generated upon the first launch of the mobile application.
This ID is a randomly generated UUID and is used only to distinguish individual installations for statistical data collection and aggregation.
Installation IDs are not linked to private user information or device characteristics.

Daily statistics include, for each installation:
\begin{enumerate}
    \item The total number of minutes of active tracking by the mobile app,
    \item The centroid of the tracked locations, approximated to a precision of ca.~0.02 latitude and longitude degrees, that is approximately 2--3~km, in order to avoid identification,
    \item The lenght of the diagonal of the bounding box containing all tracked locations,
    \item The count of known locations visited during the day,
    \item The count of notes added by the user,
    \item The count of GPS samples recorded and, possibly, discarded due to insufficient accuracy,
    \item The amount of minutes tracked at home.
\end{enumerate}

Collected information, made available as open data set, gives an indication of how the mobile app is used, allowing researchers and policy makers to gauge the effectiveness of measures adopted at regional or national level.

\subsection{Distributed queries for call to actions}

While personal data never leaves the user's device and collected statistical data cannot be used to identify users, \textit{Digital Ariadne} is designed to give designated territorial or national authorities access to the system through a dashboard allowing them to publish epidemic-related ``call to actions''.

Call to actions can be seen as geographical and temporal distributed queries that operate with the following process:
(a)~an authority creates a new call to action based on a sequence of geolocated and timestamped points and/or a set of temporary contact numbers,
(b)~the call to action is stored by the back-end service until it expires,
(c)~the mobile app automatically downloads relevant call to actions,
(d)~the mobile app matches call to actions to private location and contact data, in order to verify whether the user has been exposed to possible sources of contagion,
(e)~if there is a match, the user is privately notified and can access information associated to the call to action. Matching users may also directly choose to interact with the public authority, optionally disclosing part of their traces.

Thus, a call to action is composed of a series of geographical regions (geo-polygons), associated time intervals, and a series of temporary contact numbers~(TCNs).
The match is performed by checking whether the user has been within the indicated region in a given time period, or if any TCN is found among local records. 
Sensitivity of the match can be fine-tuned by the health authority, by indicating a maximum distance from the region and a minimum time interval of match (i.e., exposure) in order to alert the user, with the intent of reducing panic and avoiding unnecessary alarms.

\subsubsection{Creating call to actions in the case of contagion}

When diAry users are positively diagnosed, they may grant to healthcare or government authorities partial access to their local traces, inlcuding geolocations, timestamps, and the list of temporary contact numbers generated by the diAry app.

This information can be used to generate anonymous calls to action made accessible to all the instances of the app in the interested area. Call to actions are processed locally to each installation and displayed to end-users if and only if their traces match filtering criteria. This mechanism enables anonymous tracking of past interactions of diagnosed individuals, alerting potentially infected diAry users and prompting them for self-isolation and further testing.

\subsection{User incentives}

To further raise awareness and promote adoption and usage of the application, diAry integrates with the `Worth One Minute' platform.
The platform has adopted diAry as an instrument for the common good and rewards users with anonymous vouchers (called WOMs) for their collaborative behaviour~\cite{klopfenstein_worth_2019}.
These vouchers are intended to provide:
(a)~a simple \textit{gamified} experience that allows users to earn points and thus obtain positive feedback of their voluntary contribution to epidemic containment;
(b)~a tangible currency-like reward that can be adopted as a voucher system at a local and national scale to promote microeconomic growth in a post-lockdown scenario;
(c)~a perception of the social value of individual actions and behaviours.

\section{Conclusions}


In this paper, we have argued citizen empowerment to be the foundation on which novel epidemic control technologies must be built as a viable alternative to mass surveillance.
General design principles driving the development of such technologies have been outlined and applied to the design of \textit{Digital Ariadne}, an open-source privacy-preserving instrument that combines location and contact tracing capabilities to collect local traces that can be cross-matched with authoritative alerts and calls to action without leaving the end-user's device.

Just like Ariadne's thread, the data stored on personal smartphones offers a trusted trace to find a way out of the maze of COVID-19.

We invite any kind of feedback on this whitepaper, including comments on the design principles and technical contributions to the open-source diAry project.

\section*{Acknowledgments}

The authors wish to thank the more than~1000 beta testers that signed up for testing and the more than 250~users that have provided valuable feedback in the last three weeks.

\bibliographystyle{unsrt}  
\bibliography{references}

\end{multicols}

\end{document}